# PORT: A PIEZOELECTRIC OPTICAL RESONANCE TUNER

*Bin Dong[1], Hao Tian[1], Michael Zervas[2], Tobias J. Kippenberg[2], and Sunil A. Bhave[1]*
[1] OxideMEMS Lab, Purdue University, West Lafayette, IN, USA
[2] École Polytechnique Fédérale de Lausanne (EPFL), CH-1015 Lausanne, Switzerland

## ABSTRACT

This abstract presents an aluminum nitride (AlN) piezoelectric actuator for tuning optical resonance modes of silicon nitride photonic resonators. The AlN actuator is fabricated on top of a thick silicon dioxide cladding that encapsulates the nitride resonator and waveguide. The PORT is defined by undercutting the cladding layer with a lateral silicon etch. It tunes the optical wavelength by 20pm on applying 60V to the top electrode with a 0.5nA current draw. The thick oxide cladding preserves the resonator's loaded quality factor $Q_{optical}$ of 64,000 across the entire tuning range. The first bending mode is at 1.1MHz enabling a tuning speed of <1μs.

## INTRODUCTION

Optical wavelength division multiplexing (WDM) uses a laser per channel. Traditional optical WDM systems are rather large not only because of the laser but also the support electronics and thermal tuning and compensation required for each channel. Chip-scale photonic Kerr combs offer an enticing opportunity to replace a large number of lasers with just one pump laser and micro-resonator [1]. The two remaining problems in this 50Tbps datalink are comb-line alignment and drift compensation in inter-channel spacing. Currently this is achieved using heaters [2]. But even state-of-the-art coil heaters on oxide require 24mW DC power to tune 20pm and have 0.71ms response time [3].

An electrostatic silicon MEMS tuner was demonstrated in 2015 [4], [5]. But silicon is not capable of handling high optical pump power due to its two-photon absorption, precluding it from being used in Kerr comb based optical WDM. Lead zirconium titanate (PZT) has recently been used as a mechanical actuator for tuning silicon nitride resonators. However thin-film PZT has leakage at high voltages resulting in a total power consumption 0.5mW [6]. In contrast, commercially available sputtered AlN starts breaking down at 100V for a one micron thin-film and can generate substantial mechanical stress at such high input voltages. In this paper, we demonstrate tuning of 20pm with just 30nW power.

## FABRICATION PROCESS

A 187 GHz repetition rate silicon nitride comb wafer is fabricated by thick thermal oxidation followed by deposition, patterning and etching of silicon nitride waveguides and resonators. The resonators are cladded with a thick encapsulating PECVD silicon dioxide layer. More details about the fabrication process including deposition and annealing conditions can be found in [7], [8]. Then the PORT actuator consisting of Molybdenum bottom electrode, 1micron thick AlN and Aluminum top electrode is sputtered on to the wafer. After patterning the top electrode, the AlN and bottom electrodes are etched to define the piezoelectric actuator. AlN is only kept in the actuator region and nowhere else to ensure its high dielectric constant does not introduce a large RC-delay in the tuning element. AlN is deposited at 350 Celsius, so any trapped gases can bubble out of the oxide and destroy the piezoelectric film. Therefore, a high temperature anneal was performed for 10 hours to remove gases before deposition of the piezoelectric actuator layers.

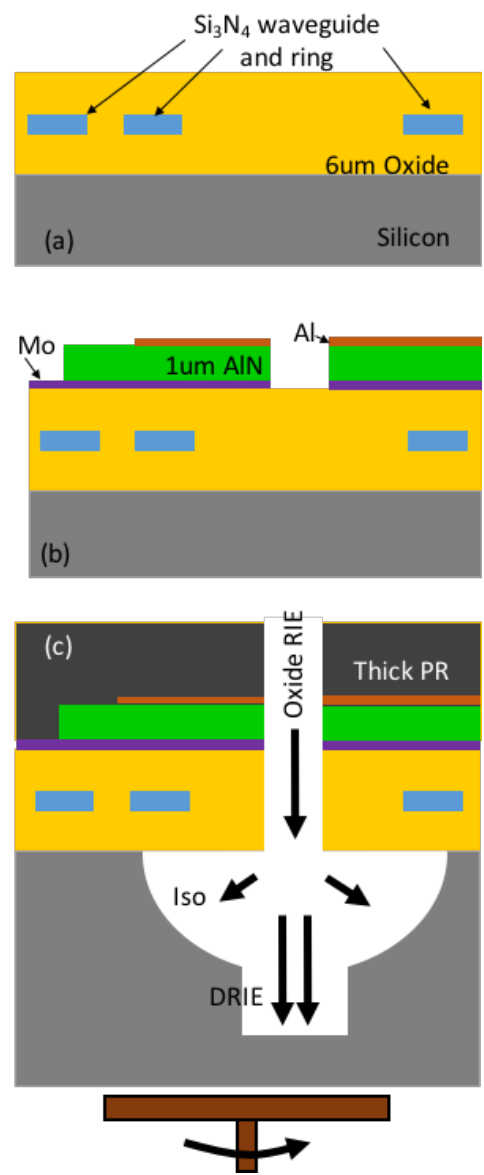

*Figure 1: (a) cross-section of the 800nm thick nitride resonator encapsulated by 6um oxide, (b) AlN actuator defined after high temp anneal to remove gases trapped in the oxide, and (c) thick PR followed by RIE oxide and silicon to release PORT, followed by backside grinding.*

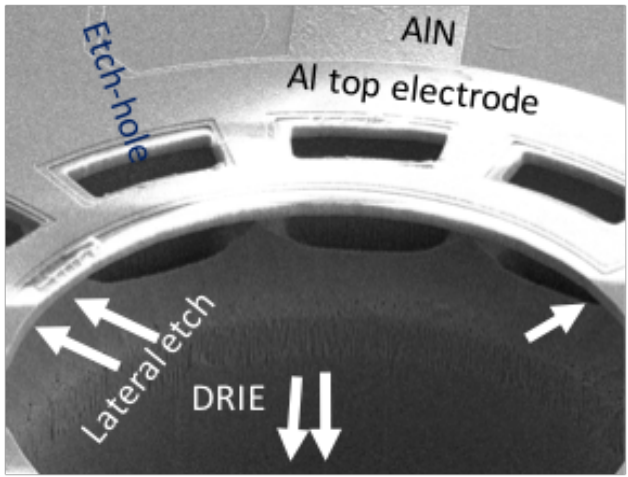
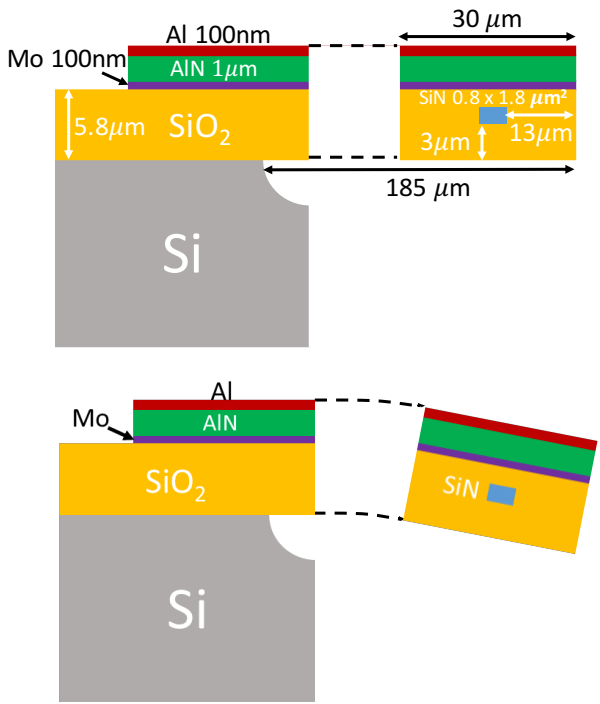

*Figure 2: (top) SEM after backside grinding. The process is almost identical to [10] except for DRIE to define facets for edge-coupling to waveguides. The total isotropic undercut of 27microns, enhanced by the etch-holes. The etch-holes are much larger than traditional MEMS processes because of the DRIE step necessary after the isotropic etch, and (bottom) Cross-section, key PORT dimensions and principle of operation.*

Once the actuator is defined, a thick photoresist mask is defined to open release windows. After an RIE etch of 6μm thick oxide an isotropic 15μm silicon etch is completed to undercut the cladding [9] and release PORT. Finally, a 200μm DRIE etch is performed to define the optical coupling facets at the chip-edge (Figure 1). After protecting the front-side, the wafer is grinded from the back to reach the DRIE trench (Figure 2), which also serves to dice the wafer. When a DC voltage is applied across the electrodes, the piezo expands (or contracts) deflecting the silicon dioxide membrane, deforming the embedded silicon nitride ring resonator.

## OPTICAL CHARACTERIZATION

The formation of an optical frequency comb depends on achieving and maintaining a very high optical $Q_{optical}$ for the entire mode family. Figure 3 compares unreleased and released resonators, showing 4× $Q_{optical}$ reduction. This is attributed to the proximity of the nitride ring to the rough surface of the oxide, as well as potential roughness and damage introduced during the subsequent silicon etch steps. However, this quality factor is still higher than silicon and thin-nitride ring resonator counterparts.

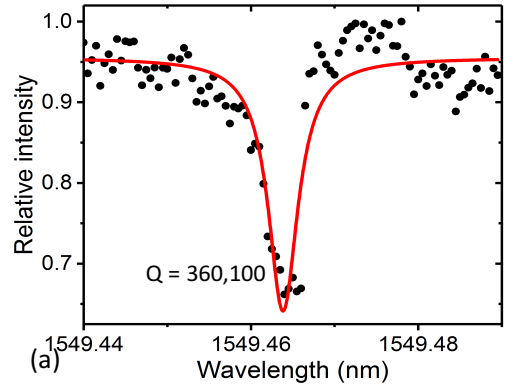
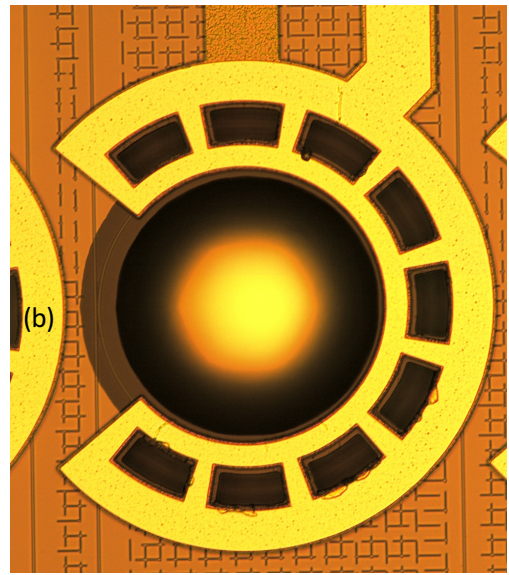
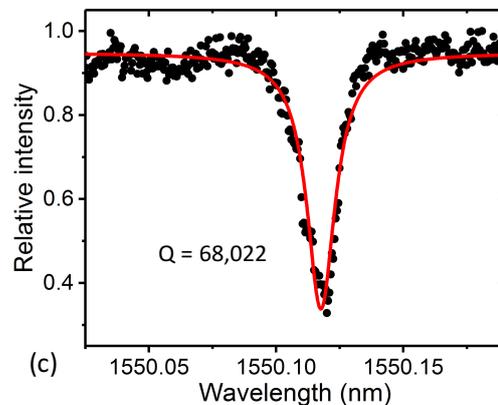

*Figure 3: (a) Optical resonance before releasing the membrane (b) microphotograph of the PORT actuator on top of the nitride resonator. The checkered pattern is the CMP fill in the nitride layer and (c) Optical resonance measurement of the released resonator.*

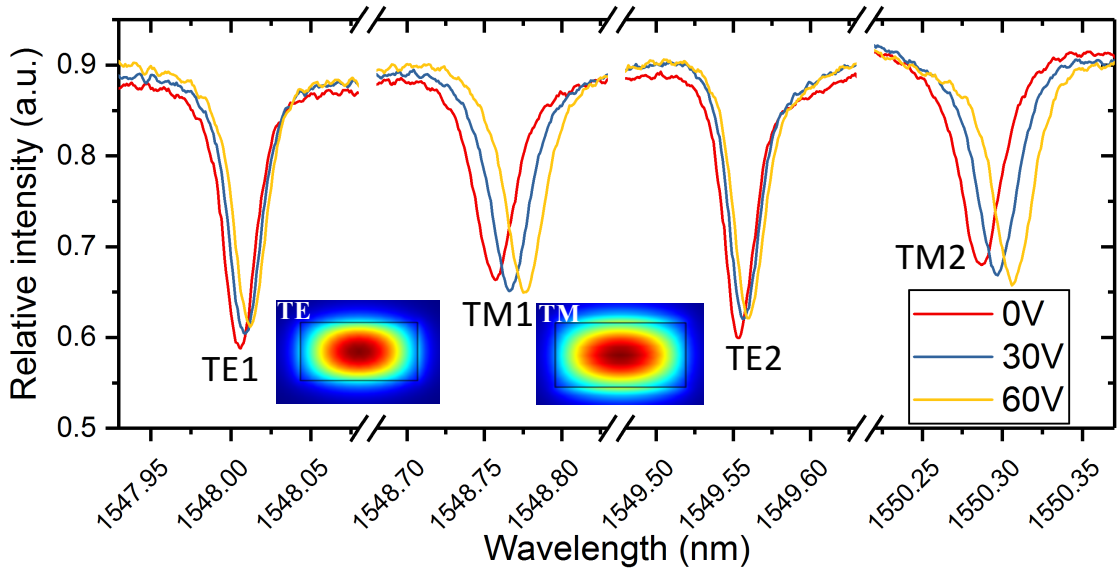

*Figure 4: PORT is operated by applying DC voltage to the top electrode. As the DC tuning voltage is ramped, the DC current draw is monitored on the Keithley source – the current never exceeded 0.5nAmps, a tuning power of 30nW.*

## PORT CHARACTERIZATION

The experimental setup is standard opto-mechanics, with pads for landing the electrical probes and inverse taper waveguides at the chip edges for optical I/O to fibers. When PORT is transduced from 0 to 60V, a red-shift in the optical transmission spectrum of 20pm if measured on the optical spectrum analyzer (Figure 4). The 20pm wavelength shift at 1550nm corresponds to 1.29% tuning. Simulations show a z-displacement of 55nm when we apply 60V to the actuator, which corresponds to a waveguide radius change of 1nm. This radius change from an original radius of 119μm is 0.84%, very close to the measured 1.29% wavelength tuning. This demonstrates PORT's tuning is not just due to geometry (shape-optic effect). Further analytical, experimental and statistical analysis of multiple devices is necessary to distill the exact value and error bound for silicon nitride's stress-optic coefficients.

When a negative voltage is applied, PORT actuates the membrane upwards generating compressive stress in the off-center positioned silicon nitride ring. If the effect was purely due to geometry change, up and down PORT motion should create identical wavelength shifts. But a negative voltage causes a blue shift (Figure 5), further confirming that silicon nitride's stress-optic component plays a significant role in tuning. Both TE and TM modes can be tuned but the TM modes are 3× more tunable than the TE modes. The optical quality factors of the TE mode are 30% higher than the TM modes, but most importantly for frequency comb formation, the $Q_{optical}$ does not change dramatically across the tuning voltage range (Figure 6).

The opto-mechanical response of PORT is shown in Figure 7. The fundamental vibration mode is the membrane flapping mode at 1.1MHz. Even though we operate PORT in air, the lack of silicon substrate right underneath the actuator minimizes the viscous damping, resulting in a mechanical quality factor of 100. Even though the potential tuning speed of PORT is 1μs, there is significant ringing from the membrane vibration mode. This directly impacts the switching speed as the membrane has to settle to a new position after every tuning step. This problem can be overcome with an overdamped design [11].

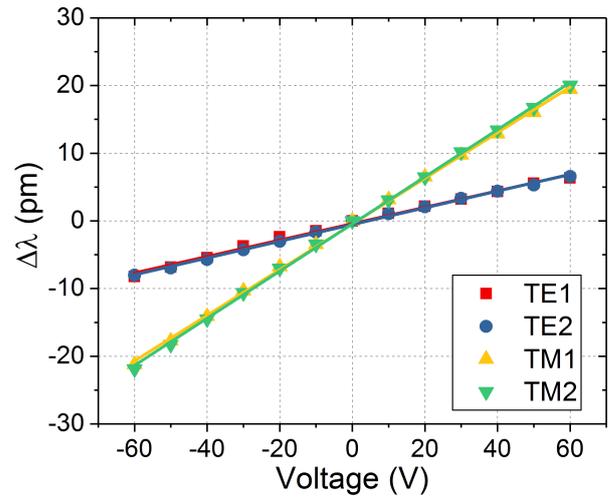

*Figure 5: The TE modes are 3× less tunable compared to the TM modes, as they have better confinement. 20pm tuning is comparable to the 3dB bandwidth of 25pm, and thus sufficient for feedback control.*

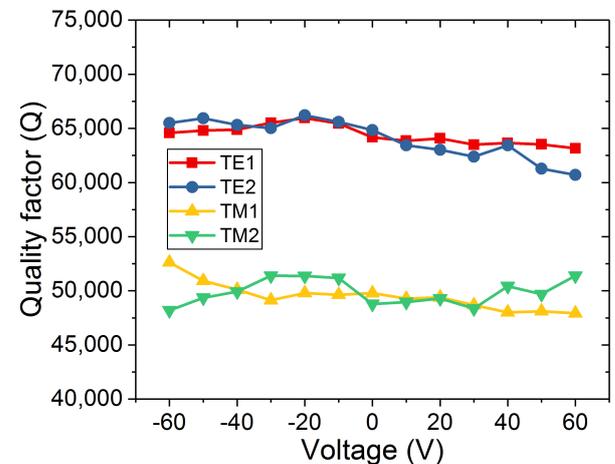

*Figure 6: The TE modes have better lateral confinement than the TM modes' vertical confinement. The quality factors do not change due to applied voltages and corresponding mechanical bending.*

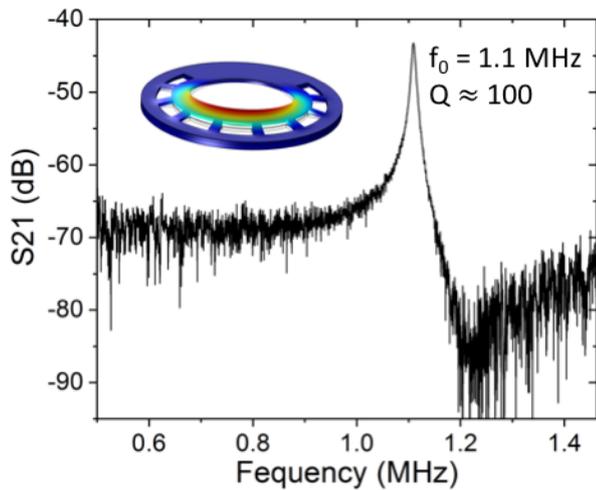

*Figure 7: Frequency response of PORT. A CW laser is slightly detuned to the optical resonance and the optical modulation due to an input RF signal is measured at the photo-detector.*

## CONCLUSIONS

We demonstrated a high-speed piezoelectric optical resonance tuner that achieves a tuning efficiency of 2pm/3nW. Unlike thermal tuning, PORT can achieve both red- and blue-tuning of the optical resonance, while maintaining constant optical quality factor across the entire tuning range. The AlN actuator is fabricated using low-temperature, CMOS-compatible process, opening the opportunity of introducing piezoelectric tuning to Electronic-Photonic integration platforms. While the PORT configuration is demonstrated as an actuator for tuning photonic resonators, an identical structure can be used for optical pickoff of mechanical sensors such as microphones and accelerometers, without the need for an ultra-small coupling gap [12].

## ACKNOWLEDGEMENTS

The authors would like to thank DARPA MTO's DODOS program and Dr. Robert Lutwak for supporting the research at Purdue University. We also want to thank M. Geiselmann and M. Karpov for discussions and initial testing.

## CONTACT

*S. Bhave, mobile:1-510-390-3269; bhave@purdue.edu